\documentclass{sf2a-conf}
\usepackage{graphicx}
\usepackage{stmaryrd}
\usepackage{amsmath}
\usepackage{url}

\newcommand{\Amp}{{\cal A}}
\newcommand{\Fbot}{F_{\hbox{\scriptsize bot}}}

\newcommand{\Tbump}{T_{\hbox{\scriptsize bump}}}
\newcommand{\Kmin}{K_{\hbox{\scriptsize min}}}
\newcommand{\Kmax}{K_{\hbox{\scriptsize max}}}

\newcommand{\na}{ \vec{\nabla} }
\newcommand{\dnz}[1]{\dfrac{d  #1}{dz}}
\newcommand{\ddnz}[1]{\dfrac{d^2  #1}{dz^2}}

\begin{document}
\TitreGlobal{SF2A 2008}
\title{Direct numerical simulations of the $\kappa$-mechanism}
\author{Gastine, T.} \address{Laboratoire d'Astrophysique de
Toulouse-Tarbes, CNRS et Universit\'e de Toulouse, 14 avenue Edouard Belin,
F-31400 Toulouse, France}
\author{Dintrans, B.$^1$} 
\runningtitle{DNS of the $\kappa$-mechanism}
\setcounter{page}{237} 

\maketitle

\begin{abstract}
We present a purely-radiative hydrodynamic model of the $\kappa$-mechanism
that sustains radial oscillations in Cepheid variables. We determine
the physical conditions favourable for the $\kappa$-mechanism to
occur by the means of a configurable hollow in the radiative
conductivity profile. By starting from these most favourable
conditions, we complete nonlinear direct numerical simulations (DNS)
and compare them with the results given by a linear-stability
analysis of radial modes. We find that well-defined instability
strips are generated by changing the location and shape of the
conductivity hollow.  For a given position in the layer, the hollow
amplitude and width stand out as the key parameters governing the
appearance of unstable modes driven by the $\kappa$-mechanism. The
DNS confirm both the growth rates and structures of the linearly-unstable
modes. The nonlinear saturation that arises is produced by intricate
couplings between the excited fundamental mode and higher damped
overtones. These couplings are measured by projecting the DNS
fields onto an acoustic subspace built from regular and adjoint
eigenvectors and a 2:1 resonance is found to be responsible for the
saturation of the $\kappa$-mechanism instability.
\end{abstract}

\section{Introduction}

Eddington (1917) discovered an excitation mechanism of stellar
oscillations that is related to the opacity in ionisation regions:
the $\kappa$-mechanism. This mechanism can only occur in regions
of a star where the opacity varies so as to block the radiative
flux during compression phases (Zhevakin 1953; Cox 1958). Ionisation
regions correspond to a strong increase in opacity, leading to the
``opacity bumps'' that are responsible for the local driving of
modes. These ionisation regions have nevertheless to be located in
a very precise region of a star, neither too close to the surface
nor to deep into the stellar core, in order to balance the damping
that occurs in other regions. It defines the so-called \textit{transition
region} which is the limit between the quasi-adiabatic interior and
the strongly non-adiabatic surface. For classical Cepheids that
pulsate on the fundamental acoustic mode, this transition region
is located at a temperature $T\simeq 40\ 000$ K corresponding to
the second helium ionisation (Baker \& Kippenhahn 1965).  However,
the bump location is not solely responsible for the acoustic
instability. A careful treatment of the $\kappa$-mechanism would
involve dynamical couplings with convection, metallicity effects
and realistic equations of state and opacity tables (Bono et al.
1999). The purpose of our model is to simplify the hydrodynamic
approach while retaining the \textit{leading order} phenomenon -the
opacity bump location- such that feasible DNS of the $\kappa$-mechanism
can be achieved.

\section{Hydrodynamic model}

\begin{figure}[]
\centering
\includegraphics[width=8cm]{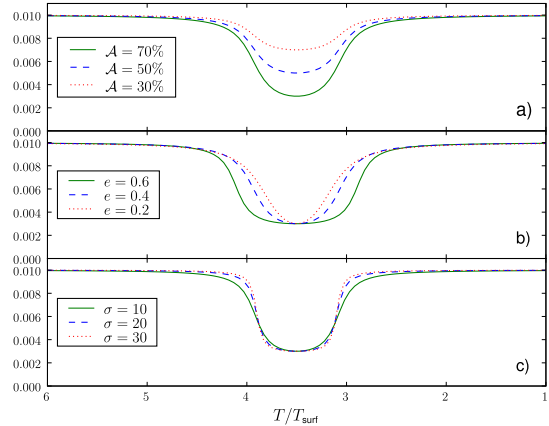}
\caption{Influence of the hollow parameters on the
conductivity profile for $\Kmax=10^{-2}$ and $\Tbump=3.5$: amplitude
$\Amp$ (\textbf{a}), width $e$ (\textbf{b}) and slope $\sigma$
(\textbf{c}).}
\label{fig:profile-rad}
\end{figure}

We focus our study on radial modes propagating in Cepheids and thus
only consider the 1-D case. Our model represents a \textit{local
zoom} about an ionisation region and is composed by a monatomic and
perfect gas ($\gamma=c_p/c_v=5/3$), with both a constant gravity
$\vec{g}$ and a constant kinematic viscosity $\nu$.  The ionisation region is
represented by a parametric conductivity hollow that mimics a bump
in opacity as (Gastine \& Dintrans 2008a):

\begin{equation}
 K_0(T)=\Kmax\left[1+\Amp\dfrac{-\pi/2+\arctan(\sigma
T^+T^-)}{\pi/2+\arctan(\sigma e^2)}\right] \text{~with~} 
 \Amp=\dfrac{\Kmax-\Kmin}{\Kmax}\text{~and~} T^{\pm}=T-\Tbump \pm e,
\end{equation}
where $\Tbump$ is the hollow position in temperature, while $\sigma$,
$e$ and $\Amp$ denote its slope, width and relative amplitude,
respectively. Examples of common values of these parameters are
provided in Fig.~\ref{fig:profile-rad}.

\section{Linear stability analysis}

We are interested in small perturbations about the hydrostatic and
radiative equilibria. The  layer is fully radiative and the radiative
flux perturbation reads under the diffusion approximation:

\begin{equation}
 \vec{F}'=-K_0\na T'-K' \na T_0,
\end{equation}
where the ``0'' subscripts mean equilibrium quantities and primes
denote Eulerian ones. The linearised perturbations obey to the
following dimensionless equations:

\begin{equation}
 \left\lbrace
 \begin{array}{rcl}
\lambda T' &=& \dfrac{\gamma}{\rho_0}\left(K_0
\ddnz{T'}+2 \dnz{K_0}\dnz{T'}+\ddnz{K_0}T'\right)
-(\gamma-1)T_0\dnz{u}+\dfrac{\Fbot}{K_0}
u,\\ \\
\lambda u &=& -\dfrac{\gamma-1}{\gamma}\left(\dnz{T'}+\dnz{\ln
\rho_0}T' +T_0
\dnz{R}\right)+\dfrac{4}{3}\nu\left(\ddnz{u}+\dnz{\ln\rho_0}
\dnz{u} \right), \\ \\\
\lambda R &=&  -\dnz{u}- \dnz{\ln \rho_0} u,
 \end{array}
 \right.
 \label{eq:syst-normal}
\end{equation}
where $R\equiv \rho'/\rho_0$ denotes the density perturbation, $u$
the velocity, $\Fbot$ the imposed bottom flux. We seek normal modes
of the form $\exp(\lambda t)$ with $\lambda=\tau+i\omega$ (unstable
modes correspond to $\tau > 0$).

\begin{figure}[htbp]
\centering
\includegraphics[width=8cm]{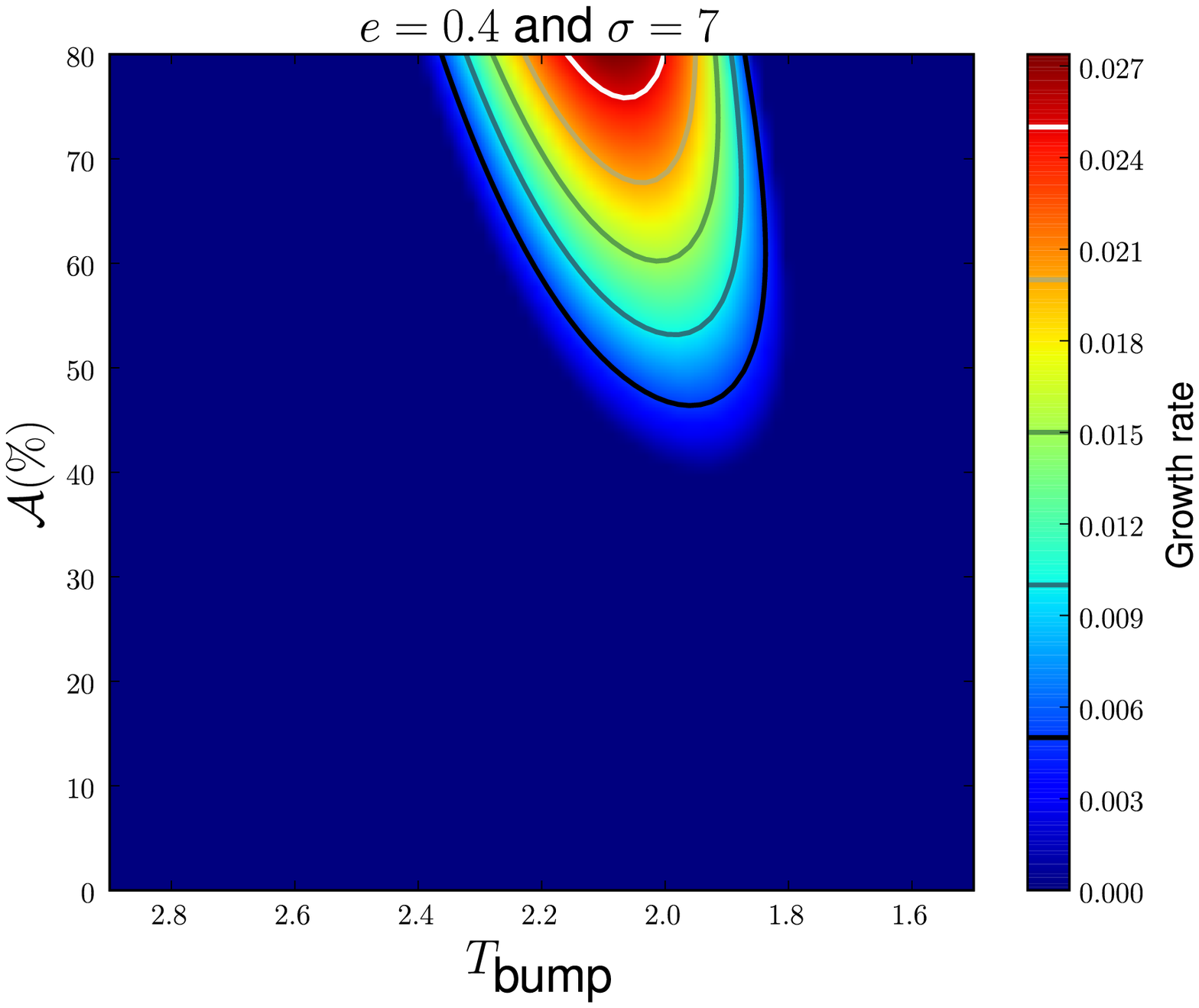}
\includegraphics[width=8cm]{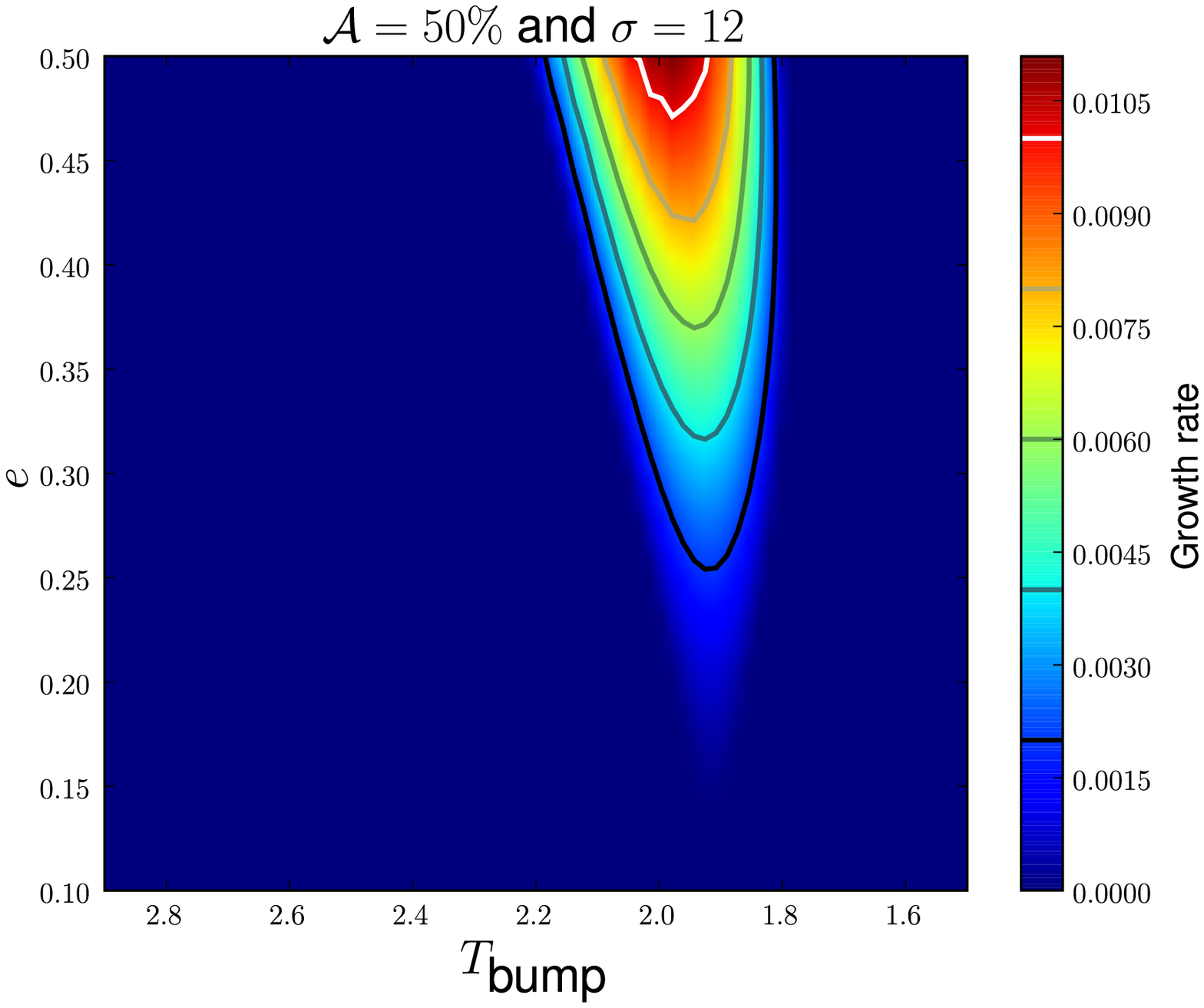}
\caption{\textit{Left panel}: Instability strip for the fundamental mode 
in the plane $(\Tbump,\ \Amp)$ for $e=0.4$, $\sigma=7$. \textit{Right
panel}: Instability strip for the fundamental mode in the plane $(\Tbump,
\ e)$ for $\Amp=50\%$, $\sigma=12$.}
\label{fig:strips}
\end{figure}

In order to investigate the influence of the hollow shape on
stability, we fix the value of $\sigma$  and vary the other parameters
($\Tbump$, $\Amp$ and $e$). For each case, we first compute the
equilibrium fields and second, the eigenvalues with their corresponding
eigenvectors are completed using the LSB spectral solver (Valdettaro
et al. 2007). Figure~\ref{fig:strips} displays the obtained instability
strips for the fundamental mode and  two main results appear: (i)
a particular region in the layer ($\Tbump\in[1.8,2.3]$) favours the
appearance of unstable modes; (ii) both a minimum width and amplitude
($e_{\text{min}}\simeq0.15$ and $\Amp_{\text{min}}\simeq 45\%$) are
needed to destabilise the system.

\section{Direct numerical simulations}

To confirm the instability strips discovered previously in the
linear-stability analysis, we perform direct numerical simulations
of the \textit{nonlinear} problem. We start from the favourable
initial conditions determined by the previous parametric surveys
and advance in time the hydrodynamic equations thanks to the
high-order finite-difference Pencil Code\footnote[2]{See
\url{http://www.nordita.org/software/pencil-code} and Brandenburg
\& Dobler (2002).}.

\begin{figure}[]
\centering
\includegraphics[width=8cm]{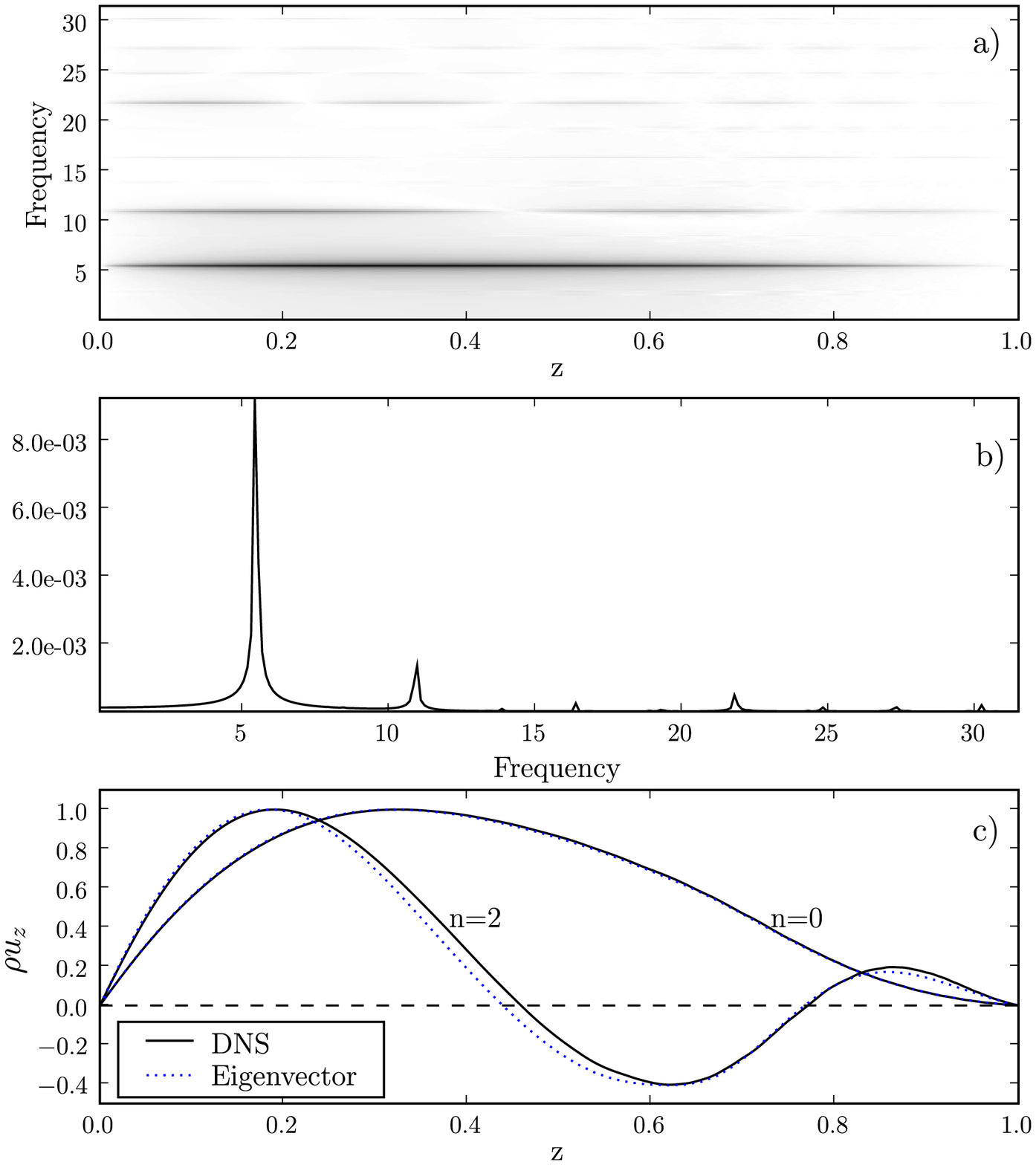}
\includegraphics[width=8cm]{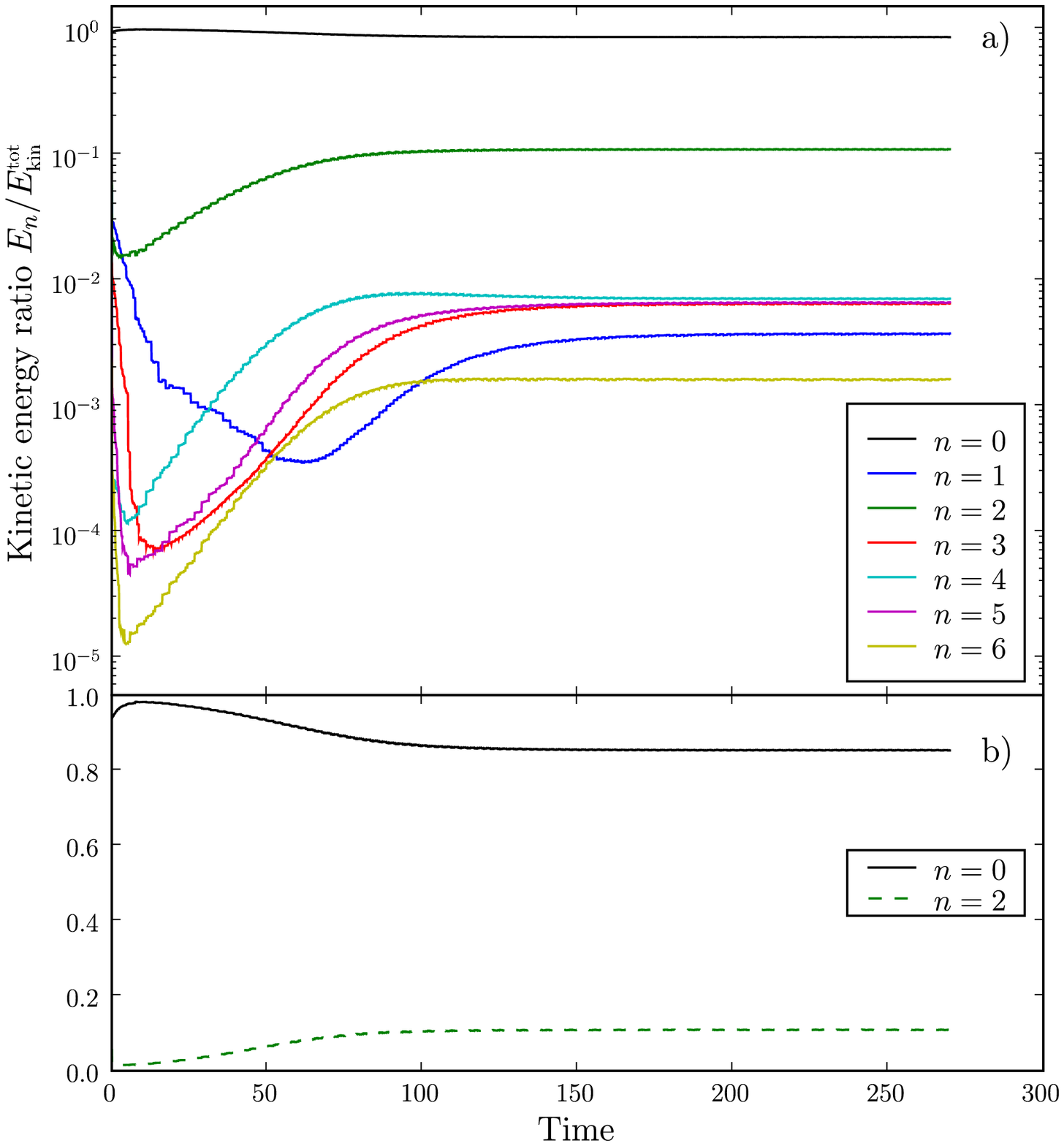}
\caption{\textit{Left panel}: a) Temporal power spectrum for the momentum
 in the $(z,\ \omega)$ plane. \textbf{b)} The resulting mean spectrum after
integrating in depth. \textbf{c)} Comparison between normalised momentum 
profiles for $n=(0,\ 2)$ modes according to the DNS power spectrum (solid
black lines) and the linear-stability analysis (dotted blue lines).
\textit{Right panel}: a) Kinetic energy ratio for $n\in\llbracket 0,\
6\rrbracket$ in a logarithmic $y$-scale. \textbf{b)} Zoom for the $n=0$
and $n=2$ modes only.}
\label{fig:fourier}
\end{figure}

To determine which modes are present in the DNS in the nonlinear-saturation
regime, we first perform a temporal Fourier transform of the momentum
field $\rho u(z,t)$ and plot the resulting power spectrum in the
$(z,\omega)$-plane (Fig.~\ref{fig:fourier}a, left). With this method, acoustic modes
are extracted because they emerge as ``shark fin profiles'' about
definite eigenfrequencies (Dintrans \& Brandenburg 2004).  We next
integrate $\widehat{\rho u}(z,\omega)$ over depth to obtain the
mean spectrum (Fig.~\ref{fig:fourier}b, left). Several discrete peaks corresponding
to normal modes appear but the fundamental mode close to $\omega_0=5.439$
clearly dominates.  Finally, the linear eigenfunctions are compared
to the mean profiles computed from a zoom taken in the DNS power
spectrum about eigenfrequencies $\omega_0=5.439$ and $\omega_2=11.06$
(Fig.~\ref{fig:fourier}c, left).The agreement between the linear-stability analysis
(eigenfunctions in dotted blue lines) and the DNS (profiles in solid
black lines) is remarkable. In summary, Fig.~\ref{fig:fourier} (left) shows that
several overtones are present in this DNS, even for long times.
However, because these overtones are linearly stable, some underlying
energy transfers must occur between modes through nonlinear couplings.

To study this nonlinear interaction, we adopt a powerful method
already used to study the sound generation by airplanes in aeroacoustics
or by compressible convection in astrophysics (Bogdan et al. 1993).
It is based on the projection of the DNS fields onto a basis shaped
from the regular and adjoint eigenvectors that are solutions to the
linear-oscillation equations.  By using projections onto these two
respective sets of eigenvectors, the time evolution of each acoustic
mode propagating in the DNS is obtained.  The kinetic energy content
of each mode is also available in this formalism, highlighting the
energy transfer between modes. As our problem only consists in an
initially static radiative zone, the velocity field that develops
is only due to acoustic modes, that is,

\begin{equation}
  E^\text{tot}_\text{kin} = E_{\text{waves}}=\sum_{n=0}^\infty
E_n,
\end{equation}
where $E_n$, is the energy contained in the $n$-acoustic mode. 

The right panel of Fig.~\ref{fig:fourier} displays the time evolution of the kinetic
energy content $E_n/E^\text{tot}_\text{kin}$ for $n \in \llbracket
0, \ 6\rrbracket$. After the linear transient growth of the fundamental
mode, a given fraction of energy is progressively transferred to
upper overtones and the nonlinear saturation is achieved above
$t\simeq 150$. These nonlinear couplings mainly involve the $n=0$
and $n=2$ modes because their energy ratios are dominant (more than
$98\%$ of the total energy). The reason for this favored coupling
lies in the period ratio existing between these two modes: the
fundamental period is $P_0=2\pi/\omega_0\simeq 1.155$, while the
$n=2$ one is $P_2\simeq 0.568$ such that the corresponding period
ratio is close to one half ($P_2/P_0\simeq 0.491$). This $n=2$ mode,
which represents about $10\%$ of the total kinetic energy, is
involved in the nonlinear saturation of the $\kappa$-mechanism
instability through a 2:1 resonance with the fundamental mode. Such
a resonance is usual in celestial mechanics with, e.g., Jupiter's
moons Io ($P=1.769$d), Europa ($P=3.551$d) and Ganymede ($P=7.154$d)
and it is well known that it helps to stabilise orbits. In our case,
this stabilisation takes the form of a nonlinear saturation: the
linear growth of the fundamental mode is balanced by the pumping
of energy from the linearly-stable second overtone behaving in that
case as an energy sink, leading to the full limit-cycle stability.

\section{Conclusion}

Direct numerical simulations (DNS) of the $\kappa$-mechanism that
excite stellar oscillations are performed. We first compute the
most favourable setups using a linear-stability analysis of radial
modes propagating in a 1-D layer of gas. In our model, a configurable
hollow in the radiative conductivity profile mimics the opacity
bump responsible for the layer destabilisation. The instability
strips found in the linear study are outstandingly confirmed by the
DNS and we show that the nonlinear saturation that arises involves
a 2:1 resonance between the linearly-unstable fundamental mode and
the linearly-stable second overtone.

\end{document}